\documentclass[conference]{IEEEtran}
\usepackage{cite}

\ifCLASSINFOpdf
  \usepackage[pdftex]{graphicx}
  \DeclareGraphicsExtensions{.pdf,.jpeg,.png}
\else
\fi
\usepackage[fleqn]{amsmath}
\usepackage{color}
\usepackage{grffile}  
\usepackage{todonotes}  
\usepackage[author={PW}]{pdfcomment}

\begin{document}
%
\title{Can Topological Transitions be Exploited to Engineer Intrinsically Quench-resistant Wires?}



%
\author{\IEEEauthorblockN{Philip Whittlesea\IEEEauthorrefmark{1},
Jorge Quintanilla\IEEEauthorrefmark{2},
James F. Annett\IEEEauthorrefmark{3}, 
Adrian D. Hillier\IEEEauthorrefmark{4} and
Chris Hooley\IEEEauthorrefmark{5}}
\IEEEauthorblockA{\IEEEauthorrefmark{1}School of Physical Sciences, University of Kent, Canterbury, Kent CT2 7NH, United Kingdom\\ Email: pw242@kent.ac.uk}
\IEEEauthorblockA{\IEEEauthorrefmark{2}School of Physical Sciences, University of Kent, Canterbury, Kent CT2 7NH, United Kingdom\\ Email: J.Quintanilla@kent.ac.uk}
\IEEEauthorblockA{\IEEEauthorrefmark{3}HH Wills Physics Laboratory, Tyndall Avenue, Bristol BS8 1TL, United Kingdom}
\IEEEauthorblockA{\IEEEauthorrefmark{4}STFC Rutherford Appleton Laboratory, Harwell Science and Innovation Campus, Oxfordshire, OX11 0QX, United Kingdom}
\IEEEauthorblockA{\IEEEauthorrefmark{5}School of Physics and Astronomy, University of St Andrews, North Haugh, St Andrews KY16 9SS, United Kingdom}}


\maketitle

\begin{abstract}
We investigate whether by synthesising superconductors that are tuned to a topological, node-reconstruction transition point we could create superconducting wires that are intrinsically resilient to quenches. Recent work shows that the exponent characterising the temperature dependence of the specific heat of a nodal superconductor is lowered over a region of the phase diagram near topological transitions where nodal lines form or reconnect. Our idea is that the resulting enhancement of the low-temperature specific heat could have potential application in the prevention of superconductor quenches. We perform numerical simulations of a simplified superconductor quench model. Results show that decreasing the specific heat exponent can prevent a quench from occurring and improve quench resilience, though in our simple model the effects are small. Further work will be necessary to establish the practical feasibility of this approach.
\end{abstract}


%
\IEEEpeerreviewmaketitle

\section{Introduction}
Quenches limit the application of superconductors in magnetic field generation, energy transmission and energy storage. When a superconductor quenches, a random fluctuation leads to a phase transition of the entire material. In short, some small finite region of the superconductor can transition into its normal state, with $T>T_{c}$, thus losing its superconducting properties. In this small normal region the superconductor suddenly has finite resistivity and will generate heat if a current is passing through it. The heat generated by this normal region transfers to the surrounding superconducting regions which can become normal as well. In turn this generates more heat until eventually the entire superconductor is heated up above $T_{c}$ and it is said to have quenched. Engineering applications of superconductors already utilise a whole host of quench prevention and protection techniques, ranging from electronic detection methods to current-sharing fail-safe systems \cite{Denz2006, Denz2001, VergaraFernandez2004, Khristi2012, Lee2009, Shajii1994} . In spite of this, quenches do occur regularly; the most well-known example being that which led to the shutdown of the LHC experiment at CERN for months in 2008 \cite{Overbye2008}.

Topology in condensed matter opens up a different avenue from the usual Ginzburg-Landau view point. Instead of changes in symmetry the focus is on changes in topology. From this view point a host of phenomena can be described, from topological insulators to Lifshitz transitions to Majorana fermions \cite{Volovik2007, Kane2005, Brumfiel2010}, none of which can be understood purely from a symmetry view point. It has been proposed that topologically protected states could be usefully applied to quantum computing \cite{Nayak2008}. Here we suggest the possibility that topological transitions in superconductors could be used to make quench-resistant wires for energy transport and storage.

Unconventional superconductors can have nodal points or lines in the quasiparticle energy spectrum where the energy gap is zero \cite{Annett1990}. A region on the Fermi surface with zero energy gap allows for arbitrarily low energy excitations. This impacts the thermodynamic properties of the system e.g. point- and line-nodes cause $T^{3}$ and $T^{2}$ specific heat dependence respectively \cite{Leggett1975, Annett1990}, rather than the exponential suppression found in conventional BCS superconductors. Recently there has been interest in anomalously-low exponents obtained at topological transitions where nodal lines cross (n=1.8), form (n=1.5) or even form and cross simultaneously (n=1.4) \cite{Mazidian2013}. These result from the non-linear dispersion of quasiparticles near the nodal regions at the transition point. 

Topological phases can occur in superconductors where different quantum ground states are identified with different topological invariants \cite{Schnyder2008}. Here we are interested in the topological node-reconstruction transition \cite{Schnyder2012, Mazidian2013} where nodal lines on the Fermi surface form, cross or reconstruct as some tuning parameter is changed, changing the associated topological number \cite{Beri2010, Mazidian2013, Fernandes2011}. In this case the associated topological number is the number of nodal lines on the Fermi surface which can only change by multiples of two.

One feature of the node-reconstruction transition is that the low temperature specific heat is enhanced, with exponents even lower than 2 or 3 \cite{Mazidian2013, Fernandes2011}. Furthermore, at finite temperature this effect is present over a swathe of tuning-parameter space; meaning it is not necessary to sit exactly at the transition to exploit the enhanced specific heat \cite{Mazidian2013}. Here we investigate whether by synthesising superconductors that are tuned to the topological, node-reconstruction transition point we could create superconducting wires that are intrinsically resilient to quenches.

Our idea is based on the principle that the superconductor in the node-reconstruction state can absorb more heat before increasing in temperature itself due to its enhanced specific heat. The more energy it can absorb before increasing in temperature the more resilient it should be to quenching. This ``passive'' approach to quench prevention is quite different to the existing approach of engineering solutions as mentioned previously. The two different approaches are entirely complementary. Materials that can be tuned to node formation/reconstruction transitions include the non-centrosymmetric superconductors Li$_{2}$(Pt,Pd)$_{3}$B \cite{Yuan2006} and the high-temperature cobalt-doped pnictides Ba(Fe$_{1-x}$T$_{x}$)$_{2}$As$_{2}$ (T=Co, Ni, Pd) \cite{Fernandes2011, Stanev2011, Mishra2009}.

\section{Model}
A cable-in-conduit-conductor, CICC, has a core made of superconducting wires with a copper matrix throughout. It is surrounded by some liquid coolant and enclosed with cladding. 

Our starting point is the general model for a CICC, found in \cite{Shajii1994}. It consists of a collection of coupled partial differential equations: a heat equation with source terms for the wire itself; another heat equation for the cladding, both depend on the temperature of the coolant; and another set of equations describing the temperature change and fluid dynamics of the coolant itself. In order to prove our concept we use a simplified version of this model which captures the relevant physics, specifically a pair of partial differential equations describing heat flow in the superconductor and in the bath:
\begin{subequations}
\label{eq:model-full}
\begin{align}
\label{eq:model-wire}
\begin{split}
  \hspace{2em}&\hspace{-2em}C^{*}_{c} ( T^{*}(x^{*}, t^{*} ), T^{*}_{c} ) \frac{\partial T^{*}(x^{*}, t^{*} ) }{\partial t^{*}} \\
  &= \frac{\partial^{2} T^{*}(x^{*}, t^{*} ) }{\partial {x^{*}}^{2}} - ( T^{*}(x^{*}, t^{*} ) - T^{*}_{h}(x^{*}, t^{*} ) ) \\
  &+ \Theta ( T^{*}(x^{*}, t^{*} ) - T^{*}_{c} )
\end{split} \\
\label{eq:model-bath}
\hspace{2em}&\hspace{-2em}\frac{\partial T^{*}_{h}(x^{*}, t^{*} ) }{\partial t^{*}} = \beta ( T^{*}(x^{*}, t^{*} ) - T^{*}_{h}(x^{*}, t^{*} ) )
 \end{align}
\end{subequations}

Here $x$ is the distance along the superconducting wire and $t$ is time. $T(x,t)$ is the temperature of the superconductor, $T_{c}$ is its critical temperature and $T_{h}(x,t)$ is the temperature of the helium bath. $C$ is the specific heat of the superconductor and $\Theta$ is a Heaviside theta function that switches on when the temperature of the conductor is above the critical temperature $T_{c}$. $\beta$ is a dimensionless constant that depends on the design of the CICC. A * indicates that a quantity has been made dimensionless by dividing by the following characteristic scales:
\begin{subequations}
\label{simple-scalars}
 \begin{align}
  x_{0} &= \left(\frac{A_{c} \kappa_{c}}{h P_{c}}\right)^\frac{1}{2} \label{x-scalar} \\
  T_{0} &= \frac{\eta_{c} I^{2} x_{0}^{2}}{A_{c}^{2} \kappa_{c}} \label{T-scalar} \\
  C_{0} &= \gamma T_{0} \label{C-scalar}\\
  t_{0} &= \frac{\rho_{c} C_{0} x_{0}^{2}}{\kappa_{c}} \label{t-scalar}\\
  \beta &= \frac{\rho_{c} C_{0} A_{c}}{\rho_{h} C_{h} A_{h}} \label{beta-scalar}
 \end{align}
\end{subequations}
Here the subscripts $c$ and $h$ differentiate between the conductor and helium respectively. $\rho$ is the density, $\kappa$ is the thermal conductivity, $h$ is the heat transfer coefficient, $P$ is the wetted perimeter and $A$ is the cross-sectional area of the conductor, $\eta$ is the resistivity, $I$ is the current in the conductor and $\gamma$ is the Sommerfeld specific heat of the superconductor.

Equation \eqref{eq:model-wire} describes how the temperature of the wire changes with time: the first term on the right hand side describes the heat flow along the length of the wire; the second term describes heat transfer with the helium bath; and the third term simulates the Joule heating that occurs when the superconductor is in the normal state.

The change in the temperature of the helium bath with respect to time, equation \eqref{eq:model-bath}, is given by a single heat transfer term which acts between the helium and the conductor - there are no terms describing the fluid dynamics of the liquid coolant. This simplification decreases computational complexity but maintains the effect of the helium increasing in temperature and helping to propagate the quench. 

\begin{figure} 
\includegraphics[width=\columnwidth]{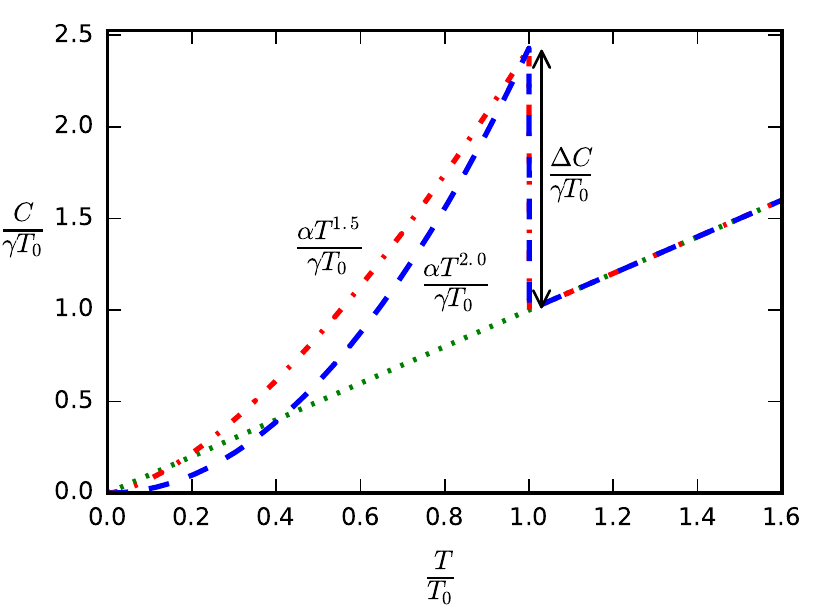}
\caption{(colour online) Specific heat of the superconductor. The green dotted line shows the linear Sommerfeld specific heat for normal metals. Our model for the nodal superconductors' specific heat is given by the blue dashed line (linear line nodes or shallow point nodes, exponent $2.0$) and the red dot-dashed line (shallow line node, exponent $1.5$). The specific heat is linear above $T_{c}$ and a power law below. At low temperatures, a lower exponent gives a higher the specific heat. The jump in specific heat at $T_{c}$ is fixed.}
\label{fig:CvModel}
\end{figure}

We model the specific heat of a nodal superconductor by 
\begin{align}
\begin{split}
 \hspace{2em}&\hspace{-2em}C^{*}_{c}(T^{*}(x^{*},t^{*}),T^{*}_{c}) \\
 &= \alpha^{*} T^{*}(x^{*}, t^{*})^{n} \Theta ( T^{*}_{c} - T^{*}(x^{*}, t^{*}) ) \\
 &+ T^{*}(x^{*}, t^{*}) \Theta ( T^{*}(x^{*},t^{*}) - T^{*}_{c} ), \label{eq:model-specificheat}
\end{split}
\end{align}
where
\begin{align}
\alpha^{*} &= \frac{\alpha}{\gamma}T_{0}^{n-1} \label{alpha-scalar}
\end{align}
and $\alpha$ and $\gamma$ are material-dependent constants. This model captures the following essential physics (see figure \ref{fig:CvModel}): \begin{itemize}\item The specific heat has a linear temperature-dependence above $T_{c}$ with some fixed Sommerfeld coefficient $\gamma$. This is appropriate for any Fermi liquid at low temperature \cite{Leggett1975}. \item At $T_c$ the specific heat has a jump, $\Delta C$, as predicted by Landau theory for any second-order phase transition \cite{Toledano1987}. For simplicity we fix the size of the jump to the value predicted by BCS theory: $\Delta C = 1.43$ \cite{de1999superconductivity}. \item Below $T_c$, the specific heat has a power law of temperature characterised by the exponent $n$ which reflects the specific nodal state \cite{Leggett1975,Annett1990,Mazidian2013}, as discussed above. The coefficient $\alpha$ is not a free parameter, but is instead fixed by the requirement that the specific heat has the right value at $T_c^-$. Its dimensionless form is $\alpha^{*} = 2.43 {T_{c}^{*}}^{1-n}$.\end{itemize}

The simplest case of this model has a constant specific heat with a constant temperature heat bath, however in this case it is easy to show that the specific heat enters only in the timescale $t_0$ of the model, i.e. the specific heat controls the rate of quench propagation and has no effect on whether a quench will happen or not. Introducing a variable specific heat while artificially maintaining a constant temperature heat bath is still not sufficient to model quench behaviour as a cold enough bath will always prevent a quench, given enough time, no matter the size or intensity of the initial heat pulse. In contrast, the model presented here contained the minimum physics required to simulate a quench. Details of the simpler models are a digression from the main point presented here, hence they have been omitted.

We assume that the system will be kept well below $T_{c}$ as this is the safest regime to prevent a quench. In this regime a nodal superconductor is preferred as its specific heat rises faster at low temperatures than that of a fully gapped superconductor. More specifically, the nodal superconductor with the lowest exponent will be preferred as it will have the highest specific heat. We can therefore study the effect different topological states can have on quench behaviour by altering the exponent $n$ in equation \eqref{eq:model-specificheat}. In this proof-of-concept work we do not consider the effect of pinning.

\section{Method and Results}
\begin{figure}
\centering
\includegraphics[width=\columnwidth]{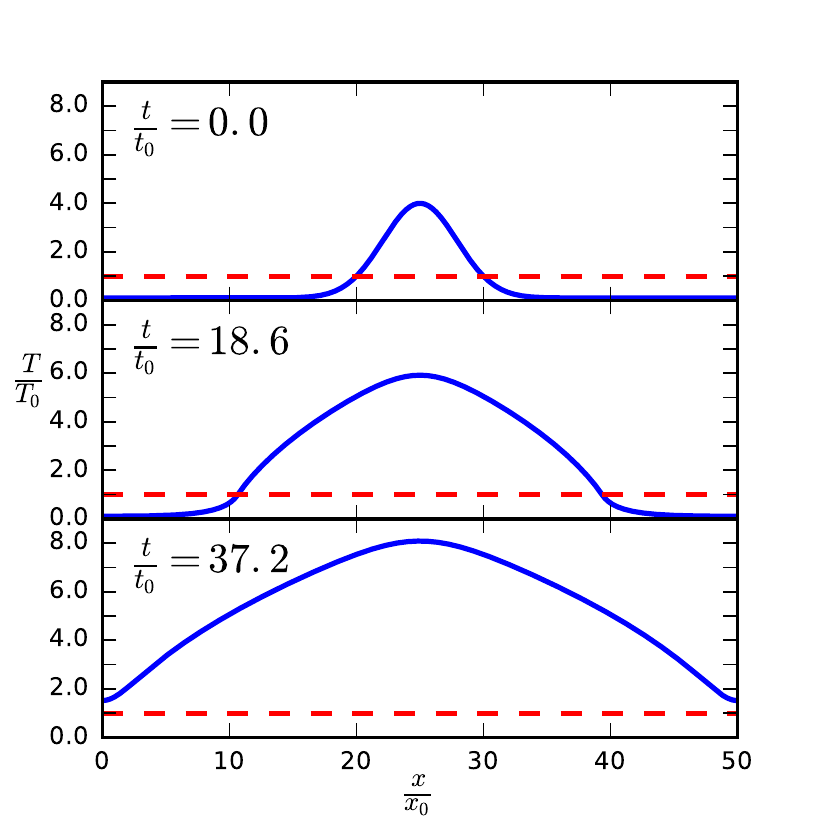}
  \caption{(colour online) Time evolution of a quench. The red dashed line is the critical temperature, the blue solid line is the temperature profile of the superconductor. Each panel shows the temperature profile at a different time, with time increasing from top to bottom. }
  \label{fig:quenchplot}
\end{figure}

\begin{figure}
\centering
\includegraphics[width=\columnwidth]{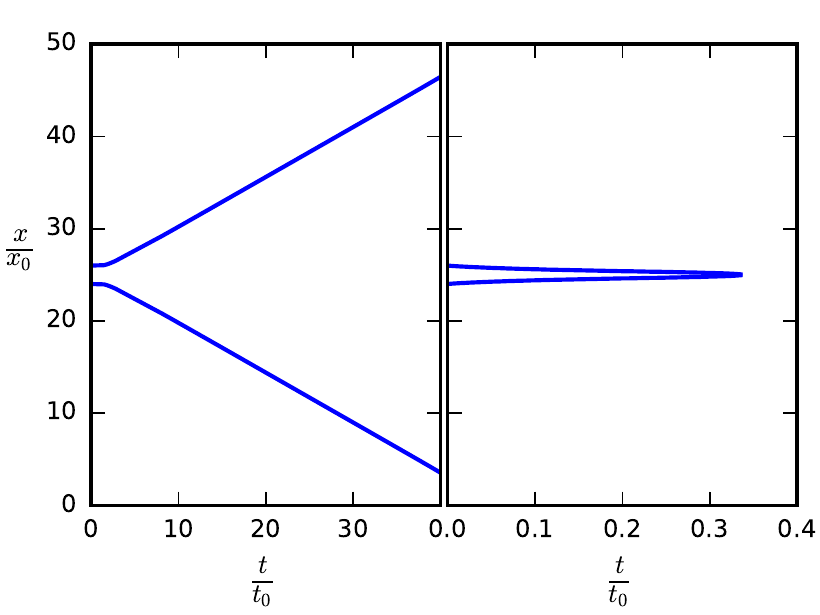}
  \caption{(colour online) Time evolution of the quench front. The solid blue line shows the position of the quench front as a function of time. The length of wire between the quench fronts is above $T_{c}$. The left panel shows a quench and the right shows a quench being prevented.}
  \label{fig:quenchfrontcomparison}
\end{figure}

The model was solved using a forward-in-time centred-in-space (FTCS) algorithm \cite{Press2007} with zero-gradient boundary conditions. The initial temperature profile of the wire has a Gaussian heat pulse centred at the middle of the wire with a temperature peak at $T_{q}$ and a width $W$ with $T > T_{c}$, called the hot zone. At the edges of the hot zone are the quench fronts: positions $x_{q}$ at which $T = T_{c}$. The quench fronts mark a boundary between the superconducting and normal regions. The time evolution of the wire's temperature profile is computed using the FTCS algorithm. A quench is said to have occurred once both the entire wire and the helium are above $T_{c}$. If the full length of the wire goes below $T_{c}$ then a quench has been prevented.

Figure \ref{fig:quenchplot} shows an example of the wire's temperature profile evolution during a quench. In this case the initial Gaussian heat pulse expands until the full length of the wire is above $T_{c}$, at which point it has quenched. The simulation continues until the helium bath is above $T_{c}$ to ensure that there is no possibility that the wire could cool down again given enough time. In the non-quenching case, the width and height of the initial Gaussian heat pulse decrease until the pulse disappears and the entire wire is below $T_{c}$.

Time evolution of the quench fronts shows the expansion or contraction of the length of the hot zone, see figure \ref{fig:quenchfrontcomparison}. If the initial conditions were correct for a quench to occur then the hot zone will increase in temperature and expand; the quench fronts will move outwards until they reach the ends of the wire. If however, the initial conditions were not sufficient for a quench, the quench fronts will move towards the centre of the wire; reducing the length of the hot zone while it decreases in temperature until the entire wire is below $T_{c}$. Additionally the quench fronts are used in determining convergence of the simulations with respect to the numerical parameters: the number of spatial divisions of the system, $dx$, the time step, $dt$ and length used to simulate an infinite wire $L$.

\begin{figure}
  \includegraphics[width=\columnwidth]{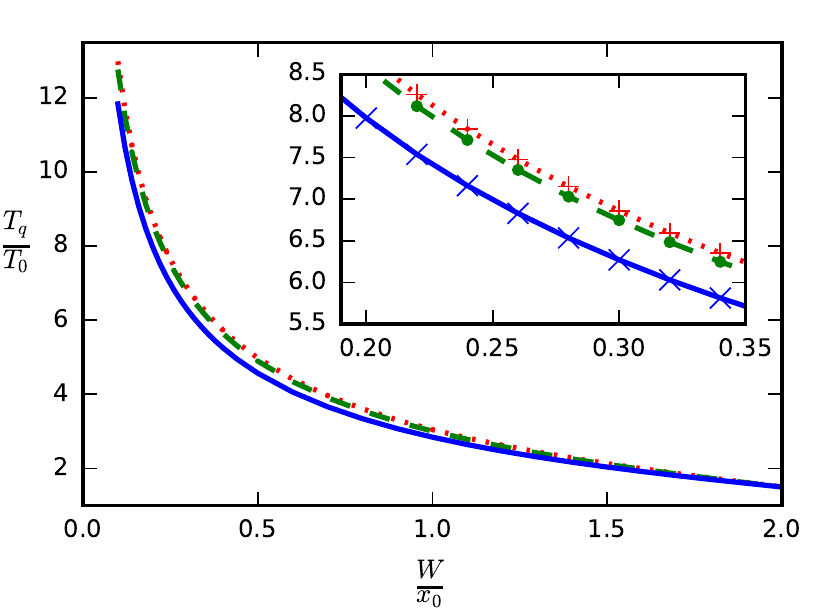}
  \caption{(colour online) Quench phase-diagram separating the parameter space into two quench and non quench regions. Each line corresponds to a different nodal state with the solid blue line characterised by $n=2.0$, the dashed green line by $n=1.5$ and the dotted red line by $n=1.4$. Here $T^{*}_{h} = 0.1 T^{*}_{c} $ but the plot stays the same for different $T^{*}_{h}$ except the width at which the lines join is $T^{*}_{h}$ dependent.}
  \label{fig:quenchphasediag}
\end{figure}

Different topological states are modelled by changing the exponent $n$ in the specific heat term in equation \protect\eqref{eq:model-specificheat}. For each topological state a `phase boundary' is constructed as a function of $W$, $T_{h}$ and $T_{q}$; see figure \ref{fig:quenchphasediag}. The parameter space is split into two regions, one which causes a quench and another which does not. All phase boundaries converge to some critical width $W_{c}$ which is $T_{h}$-dependent. For $W<W_{c}$ the phase boundaries separate and the area of parameter space that causes quenches changes with exponent. The highest exponent has the largest quench-causing area of parameter space whereas the lowest exponent ($n=1.4$ shallow line node crossing state) has the lowest quench-causing area. This smaller area in parameter space means there are fewer combinations of parameters that cause quenches, thus making the lower exponent state more resilient.

\section{Conclusion}
In this work we have investigated the effect specific heat has on the occurrence of quenches. Specifically, we have concentrated on the difference between specific heat power laws, corresponding to different nodal states with n=2 corresponding to ordinary line nodes and n=1.5, 1.4 corresponding to topological transition states \cite{Mazidian2013, Fernandes2011}. It is shown that the lower the power law exponent, the higher the specific heat and the greater the quench resilience. It is assumed that the temperature is low so that the power law approximation is valid. In this regime the power law specific heat is higher than the exponential BCS specific heat, so even the worst of the test cases would offer an improvement over BCS should the optimum topological state not be achievable.

In summary, we have asked whether a nodal superconductor could be made more resistant to quenches by tuning its parameters to a node-reconstruction topological transition point. Our calculations, using a minimum model, show that this is indeed the case although the effect is small. This concept has the potential to enhance the quench resilience of superconductors, especially if used in conjunction with current quench detection and mitigation techniques, however, detailed materials modelling will be required to ascertain whether the effect could be useful for applications. Possible candidates include the non-centrosymmetric superconductors Li$_{2}$(Pt,Pd)$_{3}$B \cite{Yuan2006} and the high-temperature cobalt-doped pnictides Ba(Fe$_{1-x}$T$_{x}$)$_{2}$As$_{2}$ (T=Co, Ni, Pd) \cite{Fernandes2011, Stanev2011, Mishra2009} This could lead to the first applications of topological transitions in the fields of energy distribution, storage and magnetic field generation.


\section*{Acknowledgment}
This research  was supported by EPSRC through the project "Unconventional Superconductors: New paradigms for new materials" (grant references EP/P00749X/1 and EP/P007392/1). JQ gratefully acknowledges a SEPnet Fellowship held during the early stages of this work. PW acknowledges a University of Kent 50th Anniversary Scholarship.

\IEEEtriggeratref{14}
\IEEEtriggercmd{\enlargethispage{-4.6in}}


\bibliographystyle{IEEEtran}
%

\end{document}